# Analyzing the Performance of Active Queue Management Algorithms

## G.F.Ali Ahammed [1], Reshma Banu[2],

[1]Department of Electronics & Communication, Ghousia college of Engg.Ramanagaram.
ali_ahammed@rediffmail.com
[2]Department of Information Science & Engg, Ghousia college of Engg.Ramanagaram.

### ABSTRACT

*Congestion is an important issue which researchers focus on in the Transmission Control Protocol (TCP) network environment. To keep the stability of the whole network, congestion control algorithms have been extensively studied. Queue management method employed by the routers is one of the important issues in the congestion control study. Active queue management (AQM) has been proposed as a router-based mechanism for early detection of congestion inside the network. In this paper we analyzed several active queue management algorithms with respect to their abilities of maintaining high resource utilization, identifying and restricting disproportionate bandwidth usage, and their deployment complexity. We compare the performance of FRED, BLUE, SFB, and CHOKe based on simulation results, using RED and Drop Tail as the evaluation baseline. The characteristics of different algorithms are also discussed and compared. Simulation is done by using Network Simulator(NS2) and the graphs are drawn using X- graph.*

### KEY WORDS

*RED; Droptail; Fairness Index; Throughput; AQM; $NS_2$ FRED, BLUE, SFB, CHOKe, ECN*

## 1. INTRODUCTION

When there are too many coming packets contending for the limited shared resources, such as the queue buffer in the router and the outgoing bandwidth, congestion may happen in the data communication. During congestion, large amounts of packet experience delay or even be dropped due to the queue overflow. Severe congestion problems result in degradation of the throughput and large packet loss rate. Congestion will also decrease efficiency and reliability of the whole network, furthermore, if at very high traffic, performance collapses completely and almost no packets are delivered.

As a result, many congestion control methods[2] are proposed to solve this problem and avoid the damage. Most of the congestion control algorithms are based on evaluating the network feedbacks[2] to detect when and where congestion occurs, and take actions to adjust the output source, such as reduce the congestion window (*cwnd*). Various feedbacks are used in the congestion detection and analysis. However, there are mainly two categories: explicit feedback and implicit feedback.

In explicit feedback algorithms, some signal packets are sent back from the congestion point to warn the source to slow down [4], while in the implicit feedback algorithms, the source deduces the congestion existence by observing the change of some network factors, such as delay, throughput difference and packet loss [4]. Researchers and the IETF proposed active queue management (AQM) as a mechanism for detecting congestion inside the network.

Further, they have strongly recommended the deployment of AQM in routers as a measure to preserve and improve WAN performance . AQM algorithms run on routers and detect incipient congestion by typically monitoring the instantaneous or average queue size. When the average queue size exceeds a certain threshold but is still less than the capacity of the queue, AQM algorithms infer congestion on the link and notify the end systems to back off by proactively dropping some of the packets arriving at a router. Alternately, instead of dropping a packet, AQM algorithms can also set a specific bit in the header of that packet and forward that packet toward the receiver after congestion has been inferred. Upon receiving that packet, the receiver in turns sets another bit in its next ACK.

When the sender receives this ACK, it reduces it transmission rate as if its packet were lost. The process of setting a specific bit in the packet header by AQM algorithms and forwarding the packet is also called marking. A packet that has this specific bit turned on is called a marked packet. End systems that experience the marked or dropped packets reduce their transmission rates to relieve congestion and prevent the queue from overflowing. In practice, most of the routers being deployed use simplistic Drop Tail algorithm, which is simple to implement with minimal computation overhead, but provides unsatisfactory performance.

To attack this problem, many queue management algorithms are proposed, such as Random Early Drop (RED) [3], Flow Random Early Drop (FRED) [4], BLUE [5], Stochastic Fair BLUE (SFB) [5], and CHOKe (CHOose and Keep for responsive flows, CHOose and Kill for unresponsive flows) [7]. Most of the algorithms claim that they can provide fair sharing among different flows without imposing too much deployment complexity. Most of the proposals focus on only one aspect of the problem (whether it is fairness, deployment complexity, or computational overhead), or fix the imperfections of previous algorithms, and their simulations setting are different from each other. These all make it difficult to evaluate, and to choose one to use under certain traffic load.

This paper aims at a thorough evaluation among these algorithms and illustrations of their characteristics by simulation. We compare the performance of FRED, BLUE, SFB, and CHOKe, using RED and Drop Tail as the evaluation baseline. For each of these algorithms, three aspects are discussed: (1) resource utilization (whether the link bandwidth is fully utilized), (2) fairness among different traffic flows (whether different flows get their fair share), and (3) implementation and deployment complexity (whether the algorithm requires too much space and computation resources).

This paper is organized as follows. In section 2, we introduce the queue management algorithms to be evaluated and how to configure their key parameters. Section 3 presents our simulation design, parameter settings, simulation result, and comparison. Section 4 discusses the key features of different algorithms, and their impact on performance. Section 5 summaries our conclusion.

## 2. QUEUE MANAGEMENT ALGORITHMS

### 2.1 RED (Random Early Drop)

RED [2] was designed with the objectives to (1) minimize packet loss and queuing delay, (2) avoid global synchronization of sources, (3) maintain high link utilization, and (4) remove biases against bursty sources. The basic idea behind RED queue management is to detect incipient congestion early and to convey congestion notification to the end-hosts, allowing them to reduce their transmission rates before queues in the network overflow and packets are dropped.

To do this, RED maintain an exponentially-weighted moving average (EWMA) of the queue length which it uses to detect congestion. When the average queue length exceeds a minimum threshold ($min_{th}$), packets are randomly dropped or marked with an explicit congestion notification (ECN) bit [2]. When the average queue length exceeds a maximum threshold ($max_{th}$), all packets are dropped or marked.

While RED is certainly an improvement over traditional drop tail queues, it has several shortcomings. One of the fundamental problems with RED is that they rely on queue length as an estimator of congestion. While the presence of a persistent queue indicates congestion, its length gives very little information as to the severity of congestion. That is, the number of competing connections sharing the link. In a busy period, a single source transmitting at a rate greater than the bottleneck link capacity can cause a queue to build up just as easily as a large number of sources can. Since the RED algorithm relies on queue lengths, it has an inherent problem in determining the severity of congestion. As a result, RED requires a wide range of parameters to operate correctly under different congestion scenarios. While RED can achieve an ideal operating point, it can only do so when it has a sufficient amount of buffer space and is correctly parameterized.

RED represents a class of queue management mechanisms that does not keep the state of each flow. That is, they put the data from the all the flows into one queue, and focus on their overall performance. It is that which originate the problems caused by non-responsive flows. To deal with that, a few congestion control algorithms have tried to separate different kind of data flows, for example Fair Queue [6], Weighted Fair Queue [6], etc. But their per-flow-scheduling philosophy is different with that of RED, which we will not discuss here.

## 2.2 FRED (Flow Random Early Drop)

Flow Random Early Drop (FRED) [4] is a modified version of RED, which uses per-active-flow accounting to make different dropping decisions for connections with different bandwidth usages. FRED only keeps track of flows that have packets in the buffer, thus the cost of FRED is proportional to the buffer size and independent of the total flow numbers (including the short-lived and idle flows). FRED can achieve the benefits of per-flow queuing and round-robin scheduling with substantially less complexity.

Some other interesting features of FRED include: (1) penalizing non-adaptive flows by imposing a maximum number of buffered packets, and surpassing their share to average per-flow buffer usage; (2) protecting fragile flows by deterministically accepting flows from low bandwidth connections; (3) providing fair sharing for large numbers of flows by using "two-packet-buffer" when buffer is used up; (4) fixing several imperfections of RED by calculate average queue length at both packet arrival and departure (which also causes more overhead).

Two parameters are introduced into FRED: $min_q$ and $max_q$, which are minimum and maximum numbers of packets that each flow is allow to buffer. In order to track the average per-active-flow buffer usage, FRED uses a global variable $avgcq$ to estimate it. It maintains the number of active flows, and for each of them, FRED maintains a count of buffer packets, $qlen$, and a count of times when the flow is not responsive ($qlen > max_q$). FRED will penalize flows with high strike values. FRED processes arriving packets using the following algorithm:

For each arriving packet P:
Calculate average queue length
Obtain connection ID of the arriving packet: flowi   connectionID(P)
if flowi has no state table then

qleni    0
strikei    0
end if
Compute the drop probability like RED: p    maxp
maxth−avg
maxth−minth
maxq    minth
if (avg _ maxth) then
maxq    2
end if
if (qleni _ maxq‖(avg _ maxth&&qleni > 2 _ avgcq)‖(qleni _ avgcq&&strikei > 1))
then
strikei    strikei + 1
Drop arriving packet and return
end if
if (minth _ avg < maxth) then
if (qleni _ max(minq, avgcq)) then
Drop packet P with a probability p like RED
end if
else if (avg < minth) then
return
else
Drop packet P
return
end if
if (qleni == 0) then
Nactive    Nactive + 1
end if
Enqueue packet P
For each departing packet P:
Calculate average queue length
if (qleni == 0) then
Nactive    Nactive − 1
Delete state table for flow i
end if
if (Nactive) then
avgcq    avg/Nactive
else
avgcq    avg
end if

**Pseudo code for the FRED algorithm**

## 2.3 BLUE

BLUE is an active queue management algorithm to manage congestion control by packet loss and link utilization history instead of queue occupancy. BLUE maintains a single probability, $P_m$, to mark (or drop) packets. If the queue is continually dropping packets due to buffer overflow, BLUE increases $P_m$, thus increasing the rate at which it sends back congestion notification or dropping packets. Conversely, if the queue becomes empty or if the link is idle, BLUE decreases its marking probability. This effectively allows BLUE to "learn" the correct rate it needs to send back congestion notification or dropping packets.

The typical parameters of BLUE are *d1*, *d2*, and *freeze_time*. *d1* determines the amount by which *Pm* is increased when the queue overflows, while *d2* determines the amount by which *Pm* is decreased when the link is idle. *freeze_time* is an important parameter that determines the minimum time interval between two successive updates of *Pm*. This allows the changes in the marking probability to take effect before the value is updated again. Based on those parameters. The basic blue algorithms can be summarized as following:

| Upon link idle event: | Upon packet loss event: |
|---|---|
| ```
if                 ((now-
last_update)>freeze_time)
     Pm = Pm-d2;
Last_update = now;
``` | ```
if                 ((now-
last_updatte)>freeze_time)
     Pm = Pm+d1;
last_update = now;
``` |

## 2.4 SFB

Based on BLUE, *Stochastic Fair Blue* (SFB) is a novel technique for protecting TCP flows against non-responsive flows. SFB is a FIFO queuing algorithm that identifies and rate-limits non-responsive flows based on accounting mechanisms similar to those used with BLUE. SFB maintains accounting bins. The bins are organized in $L$ levels with $N$ bins in each level. In addition, SFB maintains $L$ independent hash functions, each associated with one level of the accounting bins. Each hash function maps a flow into one of the accounting bins in that level. The accounting bins are used to keep track of queue occupancy statistics of packets belonging to a particular bin. As a packet arrives at the queue, it is hashed into one of the $N$ bins in each of the $L$ levels. If the number of packets mapped to a bin goes above a certain threshold (i.e., the size of the bin), the packet dropping probability $P_m$ for that bin is increased. If the number of packets in that bin drops to zero, $P_m$ is decreased. The observation is that a non-responsive flow quickly drives $P_m$ to 1 in all of the $L$ bins it is hashed into. Responsive flows may share one or two bins with non-responsive flows, however, unless the number of non-responsive flows is extremely large compared to the number of bins, a responsive flow is likely to be hashed into at least one bin that is not polluted with non-responsive flows and thus has a normal value. The decision to mark a packet is based on $P_{min}$ the minimum $P_m$ value of all bins to which the flow is mapped into. If $P_{min}$ is 1, the packet is identified as belonging to a non-responsive flow and is then rate-limited.

On every packet arrival:
Calculate hashes h0, h1, . . . , hL−1
Update bins at each level
for i = 0 to L − 1 do
if (B[i][hi].qlen > bin size) then
B[i][hi].pm   B[i][hi].pm + _
Drop packet
else if (B[i][hi].qlen == 0) then
B[i][hi].pm   B[i][hi].pm − _
end if
end for
pmin   min(B[0][h0].pm,B[1][h1].pm, . . . ,B[L − 1][hL−1].pm)
if (pmin == 1) then
ratelimit()
else
Mark or drop packet with probability pmin
end if
On every packet departure:
Calculate hashes h0, h1, . . . , hL−1
Update bins at each level
for i = 0 to L − 1 do





```
if (B[i][hi].qlen == 0) then
  B[i][hi].pm  B[i][hi].pm − _
  end if
  end for
```

**Pseudo code for the SFB algorithm**

The typical parameters of SFB algorithm are *QLen*, *Bin_Size*, *d1*, *d2*, *freeze_time*, *N*, *L*, *Boxtime*, *Hinterval*. *Bin_Size* is the buffer space of each bin. *Qlen* is the actual queue length of each bin. For each bin, *d1*, *d2* and *freeze_time* have the same meaning as that in BLUE. Besides, *N* and *L* are related to the size of the accounting bins, for the bins are organized in *L* levels with *N* bins in each level. *Boxtime* is used by penalty box of SFB as a time interval used to control how much bandwidth those non-responsive flows could take from bottleneck links. *Hinterval* is the time interval used to change hashing functions in our implementation for the double buffered moving hashing. Based on those parameters, the basic SFB queue management algorithm is shown in the above table.

## 2.5 CHOKe

As a queue management algorithm, CHOKe [4] differentially penalizes non-responsive and unfriendly flows using queue buffer occupancy information of each flow. CHOKe calculates the average occupancy of the FIFO buffer using an exponential moving average, just as RED does. It also marks two thresholds on the buffer, a minimum threshold $min_{th}$ and a maximum threshold $max_{th}$. If the average queue size is less than $min_{th}$, every arriving packet is queued into the FIFO buffer. If the aggregated arrival rate is smaller than the output link capacity, the average queue size should not build up to $min_{th}$ very often and packets are not dropped frequently. If the average queue size is greater than $max_{th}$, every arriving packet is dropped. This moves the queue occupancy back to below $max_{th}$. When the average queue size is bigger than $min_{th}$, each arriving packet is compared with a randomly selected packet, called *drop candidate packet*, from the FIFO buffer. If they have the same flow ID, they are both dropped. Otherwise, the randomly chosen packet is kept in the buffer (in the same position as before) and the arriving packet is dropped with a probability that depends on the average queue size. The drop probability is computed exactly as in RED. In particular, this means that packets are dropped with probability 1 if they arrive when the average queue size exceeds $max_{th}$. A flow chart of the algorithm is given in Figure 2. In order to bring the queue occupancy back to below $max_{th}$ as fast as possible, we still compare and drop packets from the queue when the queue size is above the $max_{th}$. CHOKe has three variants:

A. **Basice CHOKe (CHOKe):** It behaves exactly as described in the above, that is, choose one packet each time to compare with the incoming packet.
B. **Multi-drop CHOKe (M-CHOKe):** In M-CHOKe, *m* packets are chosen from the buffer to compare with the incoming packet, and drop the packets that have the same flow ID as the incoming packet. Easy to understand that choosing more than one candidate packet improves CHOKe's performance. This is especially true when there are multiple non-responsive flows; indeed, as the number of non-responsive flows increases, it is necessary to choose more drop candidate packets. Basic CHOKe is a special case of M-CHOKe with *m*=1.
C. **Adaptive CHOKe (A-CHOKe):** A more sophisticated way to do M-CHOKe is to let algorithm automatically choose the proper number of packets chosen from buffer. In A-CHOKe, it is to partition the interval between $min_{th}$ and $max_{th}$ into *k* regions, $R_1$, $R_2$, ..., $R_k$. When the average buffer occupancy is in $R_i$, *m* is automatically set as $2i$ *(i = 1, 2, ..., k).*





On every packet arrival:
if avg _ minth then
Enqueue packet
else
Draw a random packet from the router queue
if Both packets from the same flow then
Drop both packets
else if avg _ maxth then
Enqueue packet with a probability p
else
Drop packet
end if
end if

**Pseudo code for the CHOKe algorithm**

## 3. SIMULATION AND COMPARISON

In this section, we will compare the performances of FRED, BLUE, SFB and CHOKe. We use RED and Drop Tail as the evaluation baseline. Our simulation is based on ns-2 [8]. Both RED and FRED have implementation for ns-2. BLUE and SFB are originally implemented in a previous version of ns, ns-1.1, and are re-implemented in ns-2. Based on the CHOKe paper [7], we implemented CHOKe in ns-2. In our simulation, ECN support is disabled, and "marking a packet" means "dropping a packet".

### 3.1 Simulation Settings

As different algorithms have different preferences or assumptions for the network configuration and traffic pattern, one of the challenges in designing our simulation is to select a typical set of network topology and parameters (link bandwidth, RTT, and gateway buffer size), as well as load parameters (numbers of TCP and UDP flow, packet size, TCP window size, traffic patterns) as the basis for evaluation. Currently we haven't found systematic way or guidance information to design the simulation. So we make the decision by reading all related papers and extracting and combining the key characteristics from their simulations.

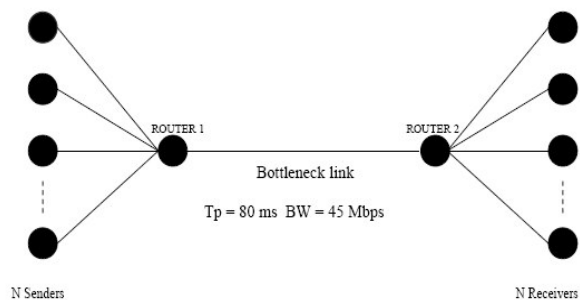

Figure 3. Simulation topology

The network topology we used is a classic dumb-bell configuration as shown in Figure 3. This is a typical scenario that different types of traffic share a bottleneck router. TCP (FTP application in particular) and UDP flows (CBR application in particular) are chosen as typical traffic patterns.





In our simulation, we use 10 TCP flows and 1 UDP flow. The bottleneck link in this scenario is the link between two gateways. We set TCP window size as 50 packets, and the router queue buffer size in the simulation as 150 packets (the packets size for both TCP and UDP are 1000 bytes). For RED, we also need to choose values for $min_{th}$ and $max_{th}$, which are typically set as 20% and 80% queue buffer size. In the following, we set them as 50 and 100 packets.

## 3.2 Metrics

*Throughput* and *queue size* are the two major metrics in our simulations. The throughput of each flow is used to illustrate the fairness among different flows, and the total throughput can be compared with the bottleneck bandwidth as an indicator of resource utilization. Queue size is a direct indicator of router resource utilization. The average queue size of each flow illustrates the fairness of router resource allocation, which also shows the different characteristics of different algorithms. We calculate the average queue size using exponentially weighted average (EWMA), and the aging weight is set to 0.002.

## 3.3 Algorithm Parameters

How to configure different algorithms for the simulation is also an issue. First, we want to show the best performance of each algorithm under the same network topology and traffic load. For the best performance, we need to fine-tune these algorithms for the fixed setting (as described above) to achieve the fairest sharing with a high utilization value. The result will be presented in Section 3.4 to show their "best-effort" performance.

On the other hand, an ideal algorithm should always achieve the best performance under all possible settings without human intervention. The different parameters set of these algorithms will impact their performance in different ways. We will discuss the impact of algorithm-specific parameters and the easiness of algorithm configuration in Section 4.

### 3.3.1 FRED

It's easy to set the parameters of FRED compared with RED. For the parameters coming from RED, FRED uses a simple formula to calculate $min_{th}$ and $max_{th}$, and assigns fixed values to $w_q$ (0.002) and $max_q$ (0.02). The only parameter new to FRED is $min_q$, whose value depends on the router buffer size. It's usually set to 2 or 4 because a TCP source sends no more than 3 packets back-to-back: two because of deployed ACK, and one more due to congestion window increase. We chose to set it to 2 (which is also the built-in setting of the FRED implementation in ns-2) after some experimentation. For most cases it turned out a FRED is not sensitive to $min_q$.

### 3.3.2 BLUE

In our simulation, the default values of BLUE static parameters are: *d1 = 0.02, d2 = 0.002, freeze_time = 0.01s*. *d1* is set significantly larger than *d2*. This is because link underutilization can occur when congestion management is either too conservative or too aggressive, but packet loss occurs only when congestion management is too conservative. By weighting heavily against packet loss, BLUE can quickly react to a substantial increase in traffic load. A rule of thrum is: *d2 = d1/10*.

### 3.3.3 SFB

The default parameter values for SFB are: *d1 = 0.005, d2 = 0.001, freeze_time = 0.001s, N=23, L=2, Boxtime = 0.05s, Hinterval = 5*. *Bin_Size* is set as *(1.5/N)* of the total buffer size of the bottleneck link. *N* and *L* are related to number of flows in the router. If the number of non-responsive flows is large while *N* and *L* are small, the TCP flows are easily misclassified as non-responsive flows [5]. Further more, since *Boxtime* indirectly determines the total bandwidth that those non-responsive flows could take in the bottleneck link, it is fine-tuned





according to different policies to treat those non-responsive flows. So in SFB, ideal parameters for one case might not necessarily good for other cases.

### 3.3.4 CHOKe

Except the parameters from RED ($min_{th}$, $max_{th}$, etc), our implementation maintains three parameters specific for CHOKe:

- `adaptive_`: control whether or not A-CHOKe should be applied, set `adaptive_ = 1` will enable A-CHOKe;
- `cand_num_`: effective when `adaptive_` is not set, when `cand_num_ = 1`, it is basic CHOKe, otherwise, it is M-CHOKe, and `cand_num_` is the number of packets to be selected from the queue;
- `interval_num_`: effective when `adaptive_` is set, and this parameters determines the number of intervals to be derided.

With our experience on running CHOKe, A-CHOKe has the best performance. So in the following simulation, we choose `adaptive_` = 1 and `interval_num_` = 5.

## 3.4 Comparison

Figure 4 and Figure 5 show the major result of the simulation. The total throughput values of all TCP and UDP flows are not shown here. For all the simulations, the total throughputs are reasonably high (about 90-96% of available bandwidth), indicating that all these algorithms provide high link utilization.

Figure 4-1 shows the UDP throughput and queue length under simulations using 10 TCP flows, 1 UDP flow, when UDP sending rate changes from 0.1Mbps to 8Mbps[1]. According to this diagram, Drop Tail is the worst in terms of unfairness, which provides no protection for adaptive flows and yields the highest UDP throughput. RED and BLUE do not work well under high UDP sending rate. When UDP sending rate is above the bottle link bandwidth, UDP flow quickly dominates the transmission on the bottleneck link, and TCP flows could only share the remaining bandwidth. On the other hand, FRED, SFB and CHOKe properly penalize UDP flow, and TCP could achieve their fair share.

One interesting point in Figure 4-1 is the behavior of CHOKe. UDP throughput decreases with the increase of UDP rate from 2Mbps to 8Mbps. This is because, with the increase of UDP rate, the total number of packets selected to compare increases, which will increase the dropping probability for UDP packets, and decrease UDP flow throughput as a result.

Figure 4-2 illustrates the size of queue buffer occupied by UDP flow. It seems that buffer usage is a good indicator of link bandwidth utilization. Similar to Figure 4-1, Drop Tail is the worst in fairness. Although RED and BLUE are similar in permissive to non-responsive flows, BLUE uses much less buffer. FRED and SFB are also the fairest.

---

[1] Due to the method for changing the UDP rate in ns-2, the sample intervals we choose are not uniform, but they will not affect our analysis





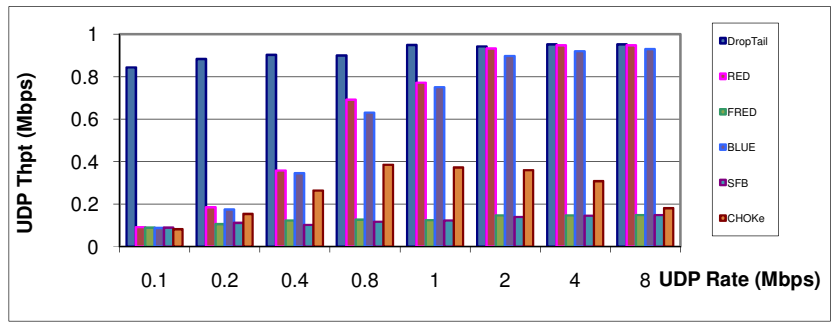

Figure 4-1.  UDP flow throughput

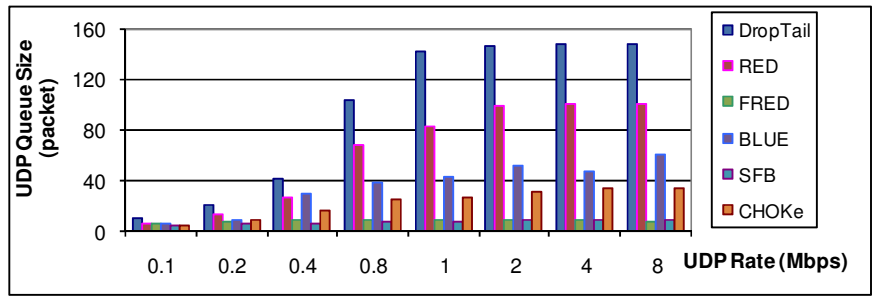

Figure 4-2. UDP flow queue size

Figure 5 illustrates the average queue size for UDP and TCP flows as well as the average total buffer usage. The difference of algorithms is clearly captured in the buffer usage plots. We can see, for Drop Tail, RED and BLUE, most of the packets in the queue are UDP flow packets, while only a small percentage belongs to TCP flows. FRED, SFB and CHOKe effectively penalize UDP flow and allow TCP flows to achieve a higher throughput.

It is also interesting to notice the difference among the total queue sizes. Since Drop Tail only drops packets when the queue buffer is full, at most time, its total queue size is the maximum queue buffer size. For RED, although it begins to provide congestion notification when the queue size reaches $min_{th}$, it only affects TCP flows, while UDP flow will keep the same sending rate, which drives the total queue size to $max_{th}$ quickly, after which all the incoming packets will be dropped and the total queue size will be kept at $max_{th}$. In CHOKe, however, the random packet selection mechanism effectively penalizes UDP flow after the average queue size reaches $min_{th}$. What's more, UDP dropping rate is proportional to its incoming rate, which will effectively keep the total queue size around $min_{th}$, as illustrated in Figure 5f. FRED, BLUE and SFB are not directly affected by $min_{th}$ and $max_{th}$ settings, so their total queue sizes have no obviously relation with these two parameters in Figure 5.

In some of the figures in Figure 5 where TCP flow queue size is very small, UDP flow queue size is the same as that of the total queue size, but the corresponding queue size for TCP flows are not zero, which seems to be a contradiction. The reason is that we draw these figures using the EWMA value of the queue size. Although we calculate the queue size every time we get a new packet, only EWMA value (weight = 0.002) is plotted[2]. It is EWMA that eliminates

---

[2] The figures of the real queue size has a lot of jitters and difficult to read.





the difference between UDP flow queue size and total queue size when TCP flow queue size is very small.

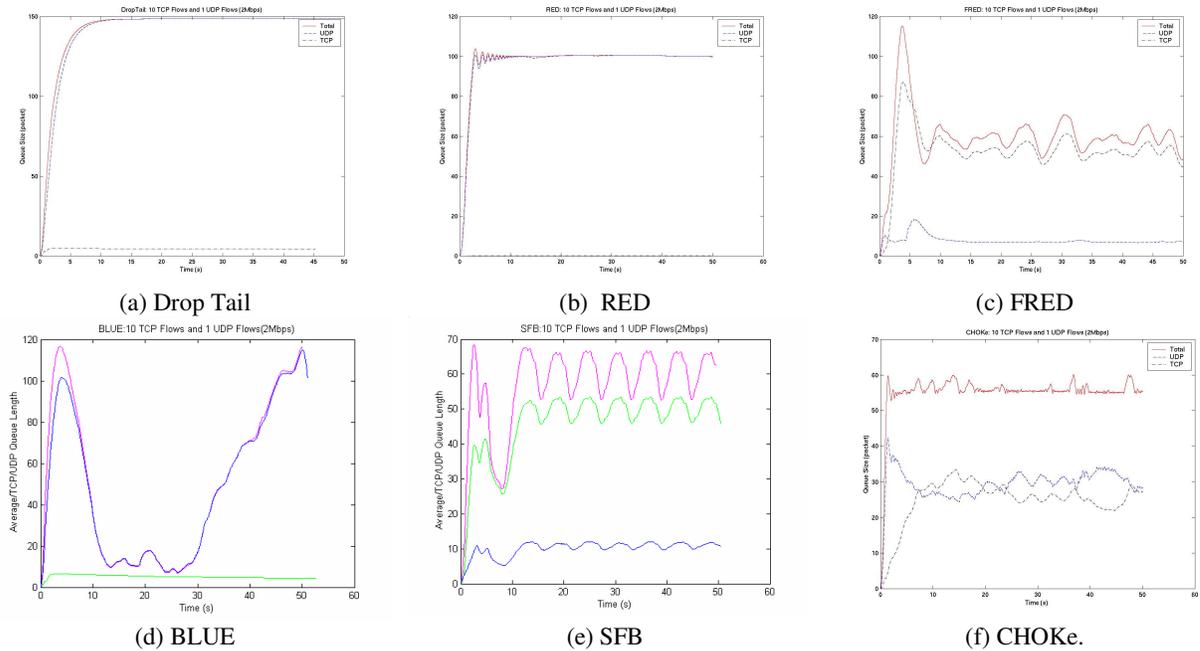

| (a) Drop Tail | (b) RED | (c) FRED |
| (d) BLUE | (e) SFB | (f) CHOKe. |

Figure 5.   Queue size in different algorithms (Notice that the total queue sizes of different algorithms are different)

# 4. ALGORITHM CHARACTERISTICS

## 4.1 FRED

FRED algorithm focuses on the management of per-flow queue length. The parameter $qlen$ is compared with $min_q$ and $max_q$, and used as a traffic classifier. Fragile flows are those whose $qlen <= min_q$, robust flows are those whose $min_q < qlen < max_q$; and non-responsive flows are those whose $qlen$ was once larger than $max_q$. The $min_q$ is set to 2 or 4, but can adapt to average queue length when there are only few robust flows (for example, in a LAN environment with small RTT and large buffer size).

FRED is very robust in identifying different kind of traffic, and protecting adaptive flows. Figure 5c shows the queue length of 1 UDP flow and the sum of 10 TCP flows. The UDP queue length was effectively limited within 10 packets, which is approximately the average queue length. The single UDP flow is isolated and panelized, without harming the adaptive TCP flows.

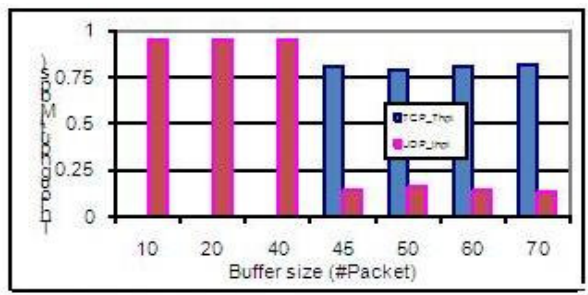





Figure 6. Impact of buffer size to FRED fairness

Figure 6 shows the impact of buffer size to FRED algorithm (without "two-packet-mode"). It is clear that FRED works well only when the buffer size is larger (larger than 45 packets in this case) enough to hold $min_q$ packets for each active flow. When the average queue length is larger than $max_{th}$, FRED degrades into Drop Tail, and can't preserve fairness. This problem is discussed in [4] and can be partly solved by using "many flow support."

The fairness of FRED is also illustrated in Table 1. The shares of UDP flows and TCP flows do not change very much as the bottleneck bandwidth increases from 0.5 Mbps to 8 Mbps. After the bandwidth of backbone link is large enough, the UDP flow gets its full share, and TCP flows begin to compete with each other.

| Bottleneck BW (Mpbs) | 0.5 | 1 | 2 | 4 | 8 | 10 | 20 |
|---|---|---|---|---|---|---|---|
| TCP_Thpt (Mbps) | 0.42 | 0.80 | 1.61 | 3.14 | 5.73 | 7.41 | 13.94 |
| UDP_Thpt (Mbps) | 0.06 | 0.15 | 0.29 | 0.66 | 1.82 | 1.85 | 1.86 |
| TCP Share (%) | 83% | 80% | 81% | 78% | 72% | 74% | 70% |
| UDP Share (%) | 12% | 15% | 14% | 17% | 23% | 19% | 9% |
| TCP Share : UDP Share | 6.92 | 5.33 | 5.79 | 4.59 | 3.13 | 3.89 | 7.78 |

Table 1. Impact of bottleneck bandwidth to FRED link utilization

The FRED algorithm has an $O(N)$ space requirement ($N$ = buffer size), which was one of the major advantages compared with per-flow queueing mechanisms (e.g. Fair Queueing). But with today's memory price, it turned out space requirement is not an important factor. The computational resources required for each packet is more significant. For each arriving packet, FRED need to classify the packet into a flow, update flow information, and calculate average queue length (also done when a packet is departing), and deciding whether to accept or drop the packet. Although optimizations can be employed to simplify the per-flow operation, it's not clear whether it can be cost-effectively implemented in backbone routers. The implementation issue is not unique to per-flow algorithms, but also applies to algorithms like RED.

In summary, FRED achieves the fairness and high link utilization by share the buffer size among active flows. It is also easy to configure, and adapt itself to preserve performance under different network environments (different bandwidth, buffer size, flow number) and traffic patterns (non-adaptive flows, robust adaptive flows, and fragile flows).

## 4.2 BLUE

The most important consequence of using BLUE is that congestion control can be performed with a minimal amount of buffer size. Other algorithms like RED need a large buffer size to attain the same goal [5]. Figure 7 shows the average and actual queue length of the bottleneck link in our simulation based on the following settings: 49 TCP flows with TCP window size 300(KB), a bottleneck link queue size 300(KB). As we can see from Figure 7, in this case the actual queue length in the bottleneck is always kept quite small (about 100KB), while the actual capacity is as large as 300KB. So only about 1/3 buffer space is used to achieve 0.93 Mbps bandwidth by TCP flows. The other 2/3 buffer space allows room for a burst of packets, removing biases against bursty sources.





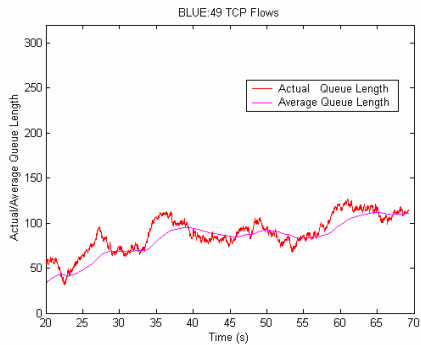
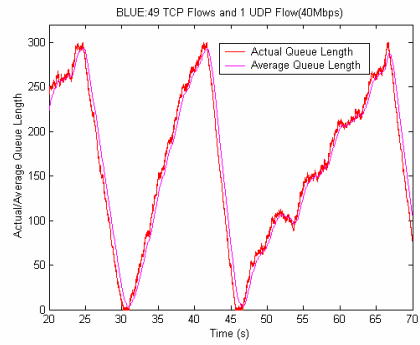

Figure 7. BLUE queue length for TCP flows          Figure 8. BLUE queue length for TCP and UDP flows

However, things get worse when non-responsive flows appear. Figure 8 show the actual and average queue length of the bottleneck link in our simulation when a 40Mbps UDP flow joins those 49 TCP flows. In this case, the total throughput (TCP and UDP) achieved is 0.94 Mbps, among which 0.01Mbps bandwidth is taken by 49 TCP flows while the UDP flow's throughput is as high as 0.93Mbps. The slow fluctuation of the bottleneck queue length shown in Figure 8   is reasonable. At $t = 40$ second, the buffer of the bottleneck link is overflowed, so $Pm$ increases to 1 quickly. Thus all the incoming packets will be dropped and in the meanwhile packets in the queue are dequeued. Since $Pm$ does not change until the link is idle, the queue length shrinks to zero gradually. The queue length at t=48s is 0. After that, the $Pm$ is decreased by BLUE. Then incoming packets could get a chance to enter queue, and the actual queue length will gradually increase from zero accordingly.

## 4.3 SFB

**(1) Basic SFB characteristics:**

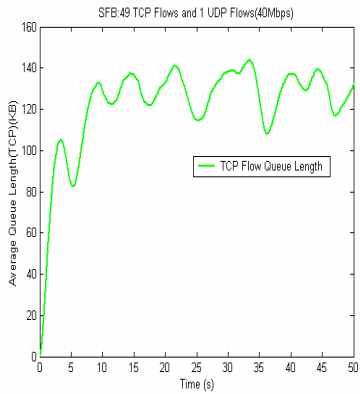
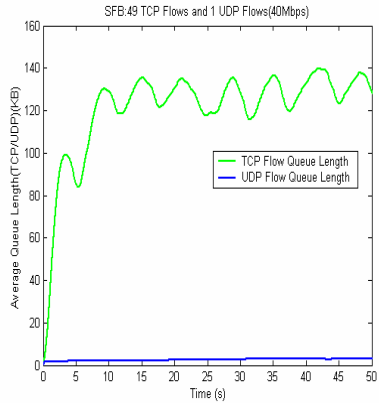

Figure 9. SFB queue length for TCP flows          Figure 10. SFB queue length for TCP and UDP flows

Although SFB is able to accurately identify and rate-limit a single non-responsive flow without impacting the performance of any of the individual TCP flows, as the number of non-responsive flows increases, the number of bins which become "polluted" or have $Pm$ values of 1 increases. Consequently, the probability that a responsive flow becomes misclassified increases. To overcome this problem, the moving hash functions was implemented, i.e. by changing the hash function, responsive TCP flows that happen to map into polluted bins will potentially be remapped into at least one unpolluted bin. However, in this case, non-





responsive flows can temporarily consume more bandwidth than their fair share. To remedy this, two set of hash functions are used to remedy this [5].

As one set of bins is being used for queue management, a second set of bins using the next set of hash functions can be warmed up. In this case, any time a flow is classified as non-responsive, it is hashed using the second set of hash functions and the marking probabilities of the corresponding bins in the warm-up set are updated. When the hash functions are switched, the bins which have been warmed up are then used. Consequently, non-responsive flows are rate-limited right from the beginning. Figure 9 and 10 show the typical performance of SFB with TCP and UDP flows. The two simulation settings are the same as that in BLUE except the buffer size of bottleneck link is set as small as 150KB.

Figure 9 shows the TCP flow queue length of the bottleneck link when there is no UDP flow. In this case, the 49 TCP flows' throughput is 0.94Mbps. While Figure 10 shows the case when a 40Mbps UDP flow joins. In this case, the UDP flow's throughput is only 0.026 Mbps while the 49 TCP flows' throughput is still quite large which consumes 0.925Mbps bandwidth of the bottleneck link. The UDP queue length is kept very small (about 4-5KB) all the time. So we could see that due to effect of SFB's double buffered moving hash, those non-responsive flows are effectively detected and rate-limited by SFB.

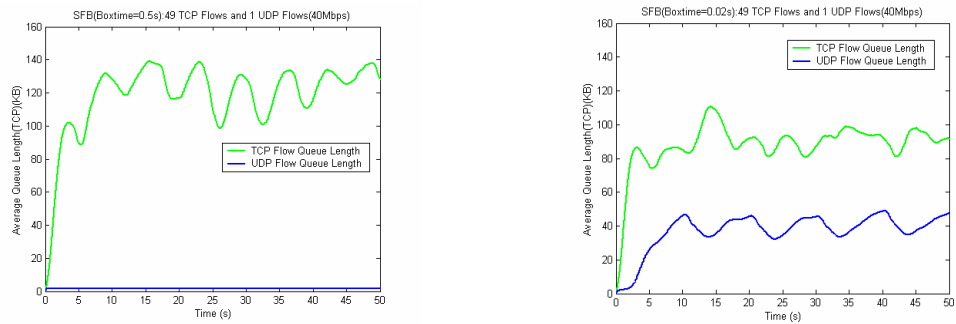

(a) Large *Boxtime*                    (b) Small *Boxtime*

Figure 11. Impact of *Boxtime* on average queue sizes of TCP and UDP flows

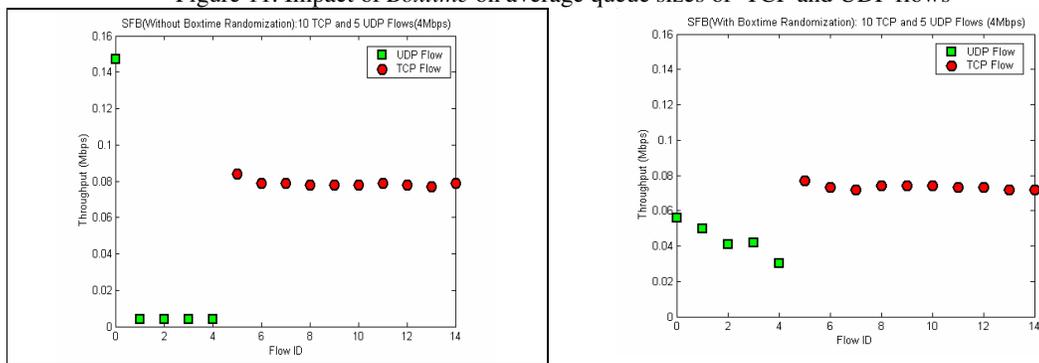

(a) Unfairness of UDP flows when rate-limited    (b) Fairness of UDP flows when rate-limited

Figure 12. Throughput of flows

## (2) Improving the fairness of UDP flows:

In SFB, all the non-responsive flows are treated as a whole. How much bandwidth those non-responsive flows could take depends mainly on the parameter *Boxtime*. The *Boxtime* is the time interval in which no packets from non-responsive flows can enter the queue. When a





packet from a UDP flow comes, if it is detected as a packet from a non-responsive flow, SFB will compare the current time with the most recent time when a packet from any non-responsive flows enqueued. If the time interval of these two events is greater than the *Boxtime*, the packet will be enqueued. Otherwise, it will be dropped. If it is enqueued, the current time is recorded for the next compare. By this way, *Boxtime* indirectly controls how much bandwidth those non-responsive flows could take. Figure 11 (a) and Figure 11 (b) illustrate the impact of *Boxtime* on the average queue size of TCP and UDP flows. In these simulations, the parameter setting is the same as the simulations illustrated by Figure 9 and Figure 10 except the value for *Boxtime*. The average queue sizes of TCP and UDP flows with a large *Boxtime* (0.5s) are illustrated in Figure 11 (a). Since large *Boxtime* means that non-responsive flows can only achieve a low throughput, the average queue length of those UDP flows is very small in Figure 11(a). In this case, the UDP flow's throughput is 0.013 Mbps, while the throughput of TCP flows is as large as 0.926Mbps. On the contrary, if the value of *Boxtime* is set small, the average queue length of those UDP flows is quite large as illustrated in Figure 11(b) when *Boxtime* is equal to 0.02 seconds. It is reasonable for the small value of *Boxtime* leads to a high throughput for those Non-responsive flows. In this case, the throughput of UDP flow is 0.27Mbps while the throughput of those TCP flows is 0.66Mbps.Since *Boxtime* is a static parameter which can only be set manually and is hard to configure automatically, the suitable value of *Boxtime* in one case might not apply to other cases. This is the main drawback of SFB.

Even worse, this rate-limited scheme can't guarantee fairness among UDP flows all the time. Figure 12(a) illustrates this case, where 5 UDP flows with sending rate 4Mbps and 10 TCP flows compete the 1Mbps bottleneck link in our standard scenario. One UDP flow (flow id 0) consumes more bandwidth than other UDP flows (flow id 1-4), although the fairness among TCP flows (flow id 5-14) is obvious. To enhance the fairness among UDP flows, the method we proposed is to make *Boxtime* a bit randomized. Figure 12(b) shows the throughput of TCP flows and UDP flows after *Boxtime* is randomized. The fairness among the UDP flows is improved. However, this method only improves the fairness of none-responsive flows when they are rate-limited to a fixed amount of bandwidth across the bottleneck. Sometimes, it is reasonable to rate-limit non-responsive flows to a fair share of the link's capacity. As suggested by [5], this could be acheived by estimating both the number of non-responsive flows and the total number of flows going through the bottleneck link.

## 4.4 CHOKe

### (1) Basic CHOKe characteristics:

Here we want to manifest the effect of CHOKe specific parameters. Our simulation setting is the same as that in section 3. That is, we use 10 TCP flows, 1 UDP flow, queue buffer size = 150 packets, $min_{th}$ = 50, $max_{th}$ = 100. Figure 13 and Figure 14 illustrate the performance of M-CHOKe and A-CHOKe with different parameters setting (`cand_num_` and `interval_num_`). From these two figures we can see increased `cand_num_/interval_num_` will effectively increase the penalization for UDP flows. This effect is most distinct when their values are small (for example, when `cand_num_` changes from 1 to 2, `interval_num_` changes from 1 to 5).





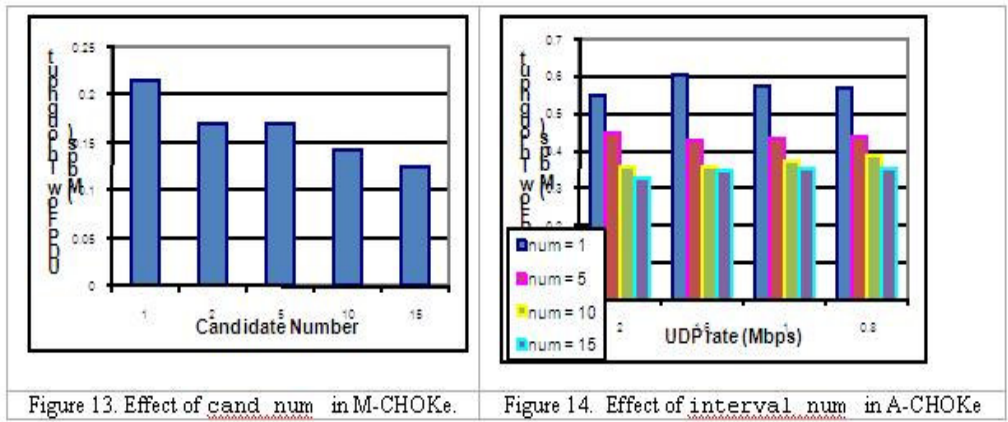

Figure 13. Effect of `cand_num` in M-CHOKe. | Figure 14. Effect of `interval_num` in A-CHOKe

## (2) Fairness of CHOKe

This simulation is to illustrate the fairness of CHOKe in managing different types of flows, comparing with RED. Simulation setting is the same as section 3. Here, we use A-CHOKe with `interval_num_` = 5. From Figure 15 we could see, when UDP rate is under the bottle link bandwidth, the performances of CHOKe and RED are very similar. When UDP rate superceeds the bottle link bandwidth, CHOKe could effectively penalize the UDP flow to keep the total queue size around $min_{th}$, and enable TCP flows to get high throughput (data not shown). Although RED keeps the queue size below $max_{th}$ most of the time, UDP rate is so high that most of queue buffer is occupied by UDP flow, which makes TCP flows throughput very low, which is the result of that RED does not distinguish different types of flows. Figure 15 tells that CHOKe could achieve much better fairness than RED.

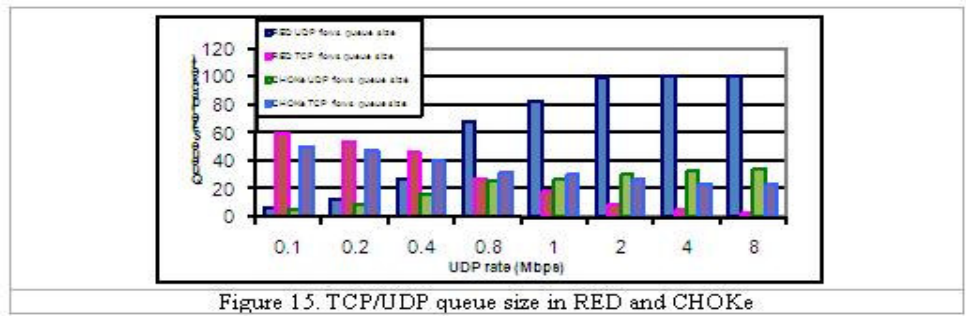

Figure 15. TCP/UDP queue size in RED and CHOKe

## (3) Effect of different number TCP and UDP flows

CHOKe can adapt to the change of the number of TCP and UDP flows. The simulation setting is the same as above, except that we change the number of TCP ($m$) and UDP ($n$) flows, with ($m$, $n$) = (1, 1), (10, 1) and (10, 5). From Figure 16a, we can see the more TCP flows, the more penalty UDP flow(s) get. It is because, with the increase of number of TCP flows, it will be less probable for the selected packet to match with the incoming TCP packet, that is, dropping of TCP packets by CHOKe decreases, correspondingly increase the throughput of TCP flows. This result could also be confirmed by the queue buffer occupation analysis (Figure 16b).

Increasing the number of UDP flows has similar effect as increasing the number of TCP flows. With more UDP flows, the percentage of UDP packets in the incoming packets will increase, which will decrease the probability to drop TCP packets. Another result is that UDP throughput decreases with increasing of UDP rate, which has been discussed in section 3.4.





On the other hand, RED will not differentiate UDP flow and TCP flow. Consequently, the throughputs have no distinct difference with the change of the number of TCP flows and UDP flows (Figure 16c, 16d).

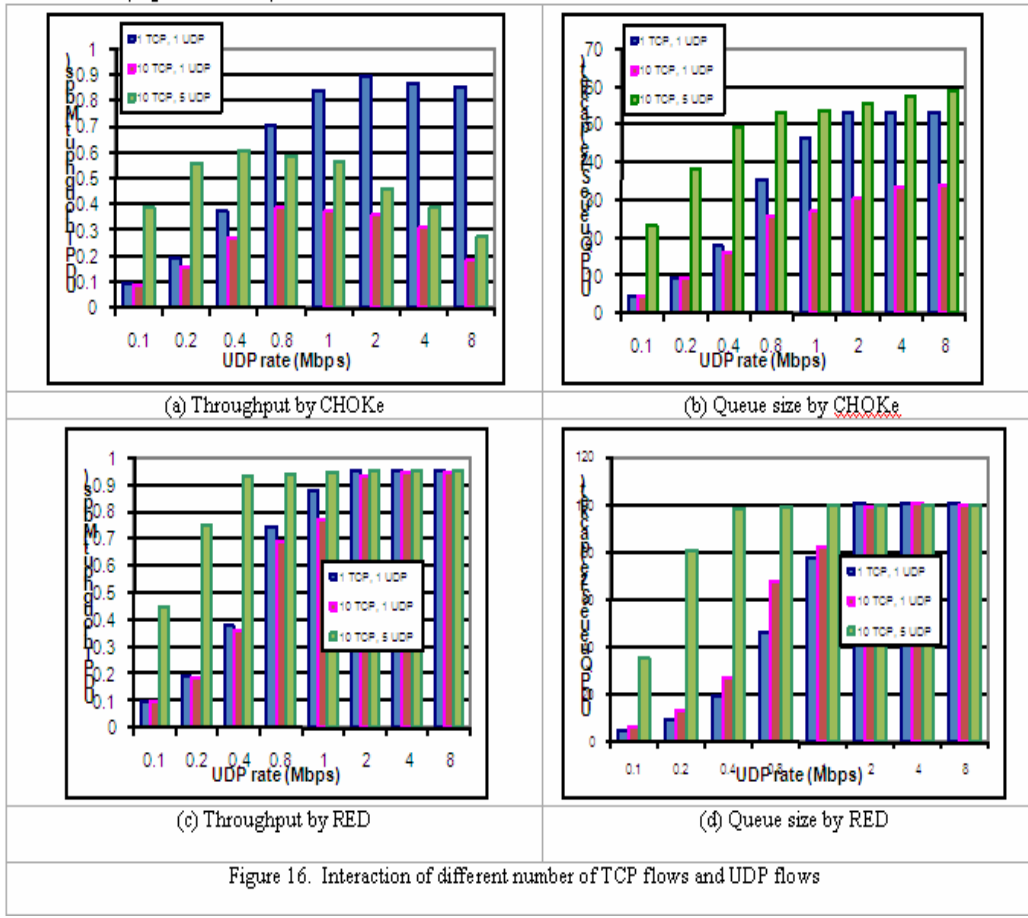

Figure 16.  Interaction of different number of TCP flows and UDP flows

**(4) RED related parameters**

We have done some simulations trying to illustrate the effect of RED related parameters on the performance of CHOKe. We find the results comply with the characteristics illustrated by the above data, which is quite predictable. For example, the throughput and queue size occupied by CHOKe changes proportionally with the setting of $min_{th}$ and $max_{th}$. And the effect of other parameters could also be explained by the analysis in [3].

# 5. CONCLUSION

This paper compared several queue management algorithms (RED, FRED, BLUE, SFB, CHOKe) based on simulation results. We have presented our simulation setting, comparison result and algorithm characteristics. It's still hard to conclude which algorithm is better in all aspects than another, especially considering the deployment complexity. But the major trends are: (1) all these algorithms provide high link utilization, (2) RED and BLUE don't identify and penalize non-responsive flow, while the other three algorithms maintains fair sharing among different traffic flows, (3) the fairness is achieved using different methods, FRED record per-active-flow information, SFB statistically multiplex buffers to bins, but needs to be reconfigured with large number of non-responsive flows, CHOKe correlates dropping rate with corresponding flow's incoming rate, and is able to penalize large number of non-





responsive flows adaptively, (4) all of the algorithms has computation overhead per incoming packet, their space requirements are different. The following table summaries our evaluation results:

| Algorithm | Link Utilization | Fairness | Space Requirement | Per-flow State Information | Configuration Complexity |
|-----------|------------------|----------|-------------------|---------------------------|--------------------------|
| RED | Good | Unfair | Large | No | Hard |
| FRED | Good | Fair | Small | Yes | Easy (Adaptive) |
| BLUE | Good | Unfair | Small | No | Easy |
| SFB | Good | Fair | Large | No | Hard |
| CHOKe | Good | Fair | Small | No | Easy (Adaptive) |

# REFERENCES


[1] Saad Biaz and Nitin Vaidya, ""De-randomizing" Congestion Losses to Improve TCP Performance over Wired-Wireless Networks" Proc. of IEEE Global Telecommun. Conf.

[2] S. Floyd and V. Jacobson, "Random early detection gateways for congestion avoidance", IEEE/ACMTransactions on Networking, vol. 1, pp 397-413, Aug, 1993.

[3] Hari Balakrishnan, Venkata Padmanabhan, Srinivasan Seshan, and Randy Katz. "Eectiveness of loss labeling in improving TCP performance in wired/wireless networks. " In Proceedings of ICNP'2002: The 10th IEEE International Conference on Network Protocols, Paris, France, November 2002.

[4] L. Brakmo and S. O'Malley, "TCP-Vegas : New techniques for congestion detection and avoidance,", In ACM SIGCOMM'94, pp. 24-35, OCT, 1994.

[5] V.Jacobson, "congestion Avoidance and Control", Proc. SIGGCOMM,pp. 314-329, Aug, 1998.

[6] R.Yavatkar and N.Bhagwat, "Improving End-to-End Performance of TCP over Mobile Internetworks" Proc of Worjshop on Mobile Computing Systems and Applications, Dec., 1994.

[7] H. balakrishnan, S. Seshan, E. Amir, R.H. Katz,

"Improving TCP/IP Performance over Wireless Networks," Proc. 1st ACM Conf. on Mobile Computing and Networking, November 1995.

[8] Sally. Floyd, Steve McCanne "Network Simulator," LBNL public domain software. Available via ftp from frp.ee.lbl.gov.

[9] Sally Floyd, "TCP and Explicit Congestion Notification," ACM Computer Communication Review, vol. 24, No. 5, October 1994.

[10] B. Bakshi, P. Krishna, N. Vaida, and D. Pradhan, "Improving performance of TCP over wireless networks," in proceedings of 17th Int. Conf. on Distributed Computing Systems. pp. 693-708, May. 1997.

[11] ISI, "ns2: network simulator" http: www.isi.edu/nsnam/ns

[12] V. Jacobson, "Modified TCP congestion avoidance algorithm," Apr 1990. mailing list, end2endinterst@isi.edu.

[13]. Dr. T. Bhaskar Reddy , Ali Ahammed ,and Reshma banu, "*Performance Comparison ofActive Queue Management Techniques*"in IJCSNS VOL.9 No.2, February 2009,pp405-408.







[14]. T. Bhaskar Reddy and Ali Ahammed"*Performance Comparison of Active Queue Management Techniques for TCP Applications* '' in JCS Journal of Computer Science 4 (12): 1020-1023, 2008


**Authors:**


**Ali Ahammed**: Received the B.Edegree in Electronics and Communication from Bangalore University ,Bangalore,India, in 2001 and M.Tech from Visvesvaraya Technological University,Belgaum, India in 2004. Currently Pursuing Ph.D in the fileld of"Computer Networks" from Sri  krishnadevaraya university ,Anantapur. India. 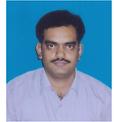

**Reshma Banu**: Received the B.Edegree in Computer Science and  Engg. from Kuvempu University, India, in 2001 and M.Tech from Visvesvaraya Technological University, Belgaum, India in 2004. Currently Pursuing Ph.D in the fileld of"Computer Networks" from Sri Krishnadevaraya university , Anantapur. India. 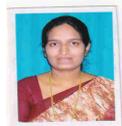